\documentclass[twoside,12pt]{article} 

\newcommand\aki[0]{\textsf{Aki}}
\newcommand\jun[0]{\textsf{Jun}}
\newcommand\rie[0]{\textsf{Rie}}
\newcommand\attr[3]{}
\def\attr#1#2.#3{\ensuremath{#1_{#2}.\mathit{#3}}}
\usepackage[dvips]{graphicx}

\begin{document} 
\pagestyle{myheadings} 
\markboth{AADEBUG 2000}{Systematic Debugging of Attribute Grammars} 
\title{ Systematic Debugging of Attribute Grammars
\footnote{In M. Ducass\'e (ed), proceedings of the Fourth International Workshop on Automated
Debugging (AADEBUG 2000), August 2000, Munich. COmputer Research Repository
(http://www.acm.org/corr/), cs.SE/0011029; whole proceedings: cs.SE/0010035.}} 
\author{ Yohei Ikezoe, Akira Sasaki, Yoshiki Ohshima, Ken Wakita,
and Masataka Sassa\\
Tokyo Institute of Technology\\
\{Yohei.Ikezoe, sasaki, ohshima, wakita, sassa\}@is.titech.ac.jp
} 
\date{} 
\maketitle 
\begin{abstract} 

Although attribute grammars are commonly used for compiler construction,
little investigation has been conducted on debugging attribute
grammars.  The paper proposes two types of systematic debugging
methods, an algorithmic debugging and slice-based debugging, both
tailored for attribute grammars.  By means of query-based interaction
with the developer, our debugging methods effectively narrow the
potential bug space in the attribute grammar description and eventually
identify the incorrect attribution rule.  We have incorporated this
technology in our visual debugging tool called \aki.

\end{abstract} 

\section {Introduction}

The \emph {attribute grammar} (AG) is a formal framework to express both syntax
and semantics of programming languages \cite {k68attributeGrammar}.
An attribute grammar description comprises a set of productions (BNF
rules) and a set of attribution rules defined over the attributes
associated with the grammar.

Attribute grammars are easy to describe and understand because they
describe ``what the programming language semantics is like'' but not
``how their attribution rules are actually implemented.''  An \emph
{AG-based compiler-compiler} takes an attribute grammar description
of a programming language and generates an efficient compiler for it.
Attribute grammar has been successfully used to describe various
programming languages and their processors.

On the other hand, debugging an attribute grammar is not simple.
Debugging an attribute grammar description using a standard debugger
exposes the attribute grammar implementation such as the attribute
evaluator, which is usually a program generated from the attribute
grammar description, and the runtime representation of attributes and
parse trees.

The paper proposes AG-aware debugging techniques for attribute grammars.
By ``AG-aware'' we mean that the debugger is aware of attribute grammars
and thus debugging does not necessarily require knowledge about strategy
and implementation of attribute evaluation.  We have formerly applied an
\emph {algorithmic debugging} \cite {s82algorithmicDebugging} technique
to attribute grammar \cite {so96aki}.  This paper is on the same line but
further incorporates \emph {slice-based debugging} technique as well.
By means of query-based interaction with the compiler developer, both
techniques effectively narrow the potential bug space in the attribute
grammar description and eventually identify the incorrect attribution
rule.

The benefit of our approach, independence from the implementation
of particular AG-based compiler-compiler, is twofold:  (1) the
compiler developer is freed from understanding implementation of the
compiler-compiler and (2) the proposed technique can be applied to other
AG-based system.

Our hybrid debugging technique has been implemented as a visual debugging
environment called \aki.  It is written in Squeak Smalltalk \cite {di97Squeak}.
The resulting system has been used in our project that
incorporates AG technologies in most phases of compiler construction ---
including transformation of intermediate code, optimization, and code
generation \cite {s91RieJun}.

The rest of this paper is as follows.  Section~\ref {sec: attribute
grammar} briefly introduces the attribute grammar and attribute
evaluation, sections~\ref {sec: algorithmic debugging} and~\ref
{sec: slice-based debugging} explain the two debugging techniques,
section~\ref {sec: aki} describes the visual debugger, 
section~\ref {sec: discussion} discusses the debugger,
and section~\ref {sec: conclusion} concludes this article.

\section{Attribute grammars}
\label{sec: attribute grammar}

  The attribute grammar \cite{k68attributeGrammar} is a description that
provides both syntax and semantics of an input string.  It is one of
fundamental system used for formalization and construction of compilers.

\renewcommand\baselinestretch{1.0}
\begin{figure}

\begin{tabbing}
 F \ \ \=::= \ \ \= . \ \ L \\
        \> \{ \attr L{}.{pos} = 1; \\
        \> \ \ \attr F{}.{val} = \attr L{}.{val} \} \\
L$_0$   \>::=     \> B \ \ L$_1$ \\
        \> \{ \attr L1.{pos} = \attr L0.{pos} + 1; \\
        \> \ \ \attr B{}.{pos} = \attr L0.{pos} + 1; \ \ \ \ \ \ \ \ (bug) \\
        \> \ \ \attr L0.{val} = \attr B{}.{val} + \attr L1.{val} \} \\
        \> \ \ $\mid$ \> B \\ 
        \> \{ \attr B{}.{pos} = \attr L0.{pos}; \\
        \> \ \ \attr L0.{val} = \attr B{}.{val} \} \\
B     \>::=     \> 1 \\
        \> \{ \attr B{}.{val} = $2^{-\attr B{}.{pos}}$ \} \\
        \> \ \ $\mid$ \> 0 \\
        \> \{ \attr B{}.{val} = 0 \} 

\end{tabbing}
  \caption{An example of attribute grammar}
  \label{fig: grammar}
\end{figure}
\renewcommand\baselinestretch{1.0}

\begin{figure}
 \centerline{\includegraphics[width=0.7\linewidth]{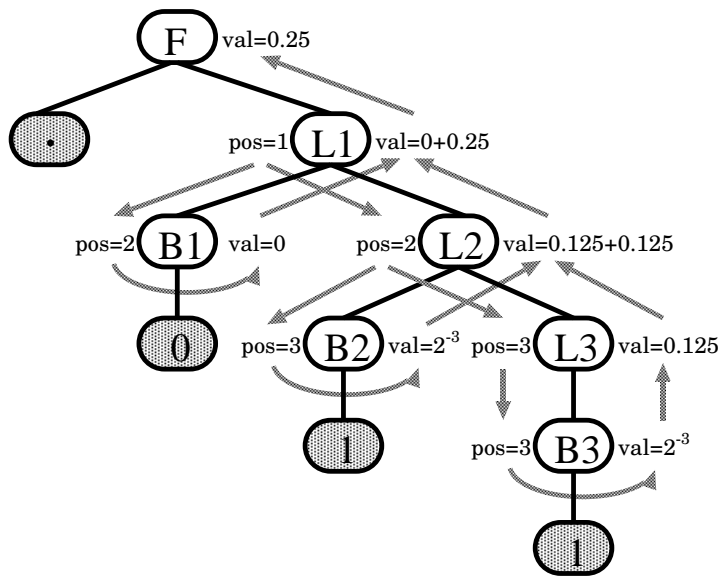}}
 \caption{Attributed parse tree}
 \label{fig: parse tree}
\end{figure}

  Fig.~\ref{fig: grammar} is an example of an AG description that
evaluates binary representation of numbers.  An attribute evaluator
generated from the AG description takes an input string (e.g., ``.011'')
and parses it in accordance with the syntax definition of the grammar.
Then the attribute evaluator computes attribute values for each node in
the parse tree according to the dependency among attributes 
and associates them to the respective node.  In this manner,
an \emph {attributed parse tree} is created from the input string (see
Fig.~\ref {fig: parse tree}).  The shaded arrows represent dependency
between attributes.

  There are two kinds of attributes.  One is called {\it inherited}
attribute, whose value is computed from the values of the attributes
associated with the ancestor and sibling nodes.  The other is called
{\it synthesized} attribute, whose value is computed from the values of
the attributes associated with the children nodes.

  We have intentionally introduced a bug in the AG description for the
sake of discussion in the following sections.  Because of this, some of
the attribute values in Fig.~\ref {fig: parse tree} are wrong.

\section{Application of algorithmic debugging to attribute grammar}
\label {sec: algorithmic debugging}

Algorithmic debugging \cite {s82algorithmicDebugging} is a systematic
bug locating technique.  With programmer's guidance, an algorithmic
debugger locates a bug in program execution.  Algorithmic debugging
formulates execution in terms of \emph {computation tree} which is defined
as recursive logical deduction of logic programming, or recursive
$\beta$-reduction for functional programming.  The debugging method
has been applied to functional \cite {Nilsson94} or procedural language
\cite {f92GADT}.

To apply algorithmic debugging to attribute grammar paradigm, we need to
formulate attribute evaluation as some form of recursive application.
In \cite {so96aki} we have shown that the notion of \emph {Synth function}
\cite {m81AGM} suits for this purpose.

It is known that for any synthesized attribute $s$ of a node $N$
in the parse tree, its value $N.s$ is uniquely defined using the Synth
function $F_{N.s}$:
\[ N.s = F_{N.s} (N.I_{N.s}, \mathit {tree}_N). \]
where $\mathit {tree}_N$ stands for the subtree of the parse tree rooted
at node $N$ and $N.I_{N.s}$ for a set of inherited attributes
of $N$ on which $N.s$ directly or indirectly depends in $\mathit {tree}_N$
\cite {so96aki}.

Because the entire attribution for the parse tree can be represented as
recursive application of Synth functions, we can use Synth functions as
basis for formulating computation tree for attribute grammars.

\begin{figure}
 \centerline {\includegraphics[width=0.7\linewidth]{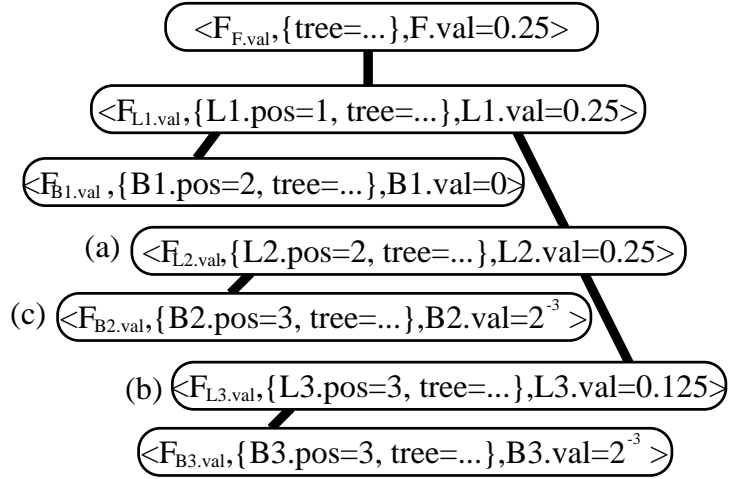}}
 \caption{Computation tree}
 \label{fig: computation tree}
\end{figure}

  We will explain how the bug in Fig.~\ref{fig: grammar} can be detected
by the algorithmic debugging.  When a string ``.011'' is given as an input
to the description, the computation tree in Fig.~\ref{fig: computation
tree} is created.  First the debugger chooses node (a) in this figure, and
queries the user whether $\attr
L2.{val} = 0.25$ is correct with respect to the argument \attr L2.{pos} and
the subtree rooted at $L_2$.  This query means whether the binary number
``11'' from the second decimal place represents 0.25, and the user
can respond that this is incorrect because the correct value is 0.375.
Given this, the debugger prunes the computation tree above (a) from the
search space.  Then the debugger queries similar question for node (b).
This time the value is correct and the subtree rooted at (b) is pruned.
Next the debugger queries similar question for node (c).
Finally the debugger locates the attribution rule for node (a) as containing
a bug.

  As for the applicability of the method to classes of AG, 
\cite {so96aki} gives how to apply algorithmic debugging to noncircular AG, 
absolutely noncircular AG, simple multi-visit AG and to its subclasses.

\section {Slice-based debugging}
\label {sec: slice-based debugging}

Techniques that utilize the notion of \emph {program slices} have been
applied to program verification, program testing, version management,
and systematic program debugging.
Shimomura proposed a systematic debugging method of procedural programming
languages using slices \cite {s93CriticalSlice}.  We applied this idea 
to the attribute grammar framework.

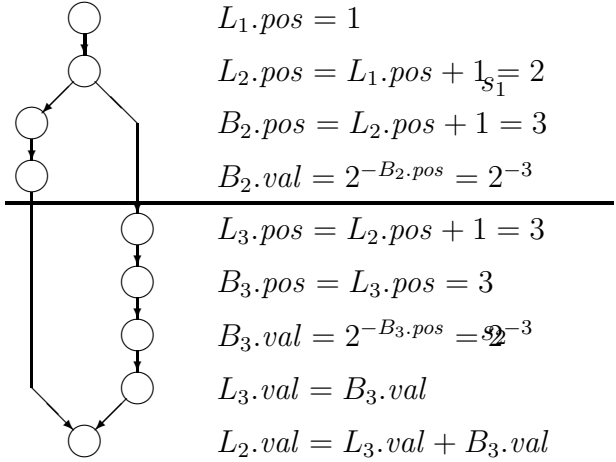
\begin {figure}
\begin {center}

\newcommand\Circle[3]{\put(#1,#25){\circle{6}}\put(40,#22){\makebox(0,0)[lb]{$#3$}}}
\def\lCircle(#1,#2){\Circle{5}{#1}{#2}}
\def\cCircle(#1,#2){\Circle{15}{#1}{#2}}
\def\rCircle(#1,#2){\Circle{25}{#1}{#2}}

\begin{tabular}{lr}
\begin{minipage}{0.42\textwidth}
\setlength{\unitlength}{2pt}
\begin {picture}(100,100)(0,0)
\linethickness {1pt}
\cCircle(9,\attr L1.{pos} = 1)
  \put(15,92){\vector(0,-1){4}}
\cCircle(8,\attr L2.{pos} = \attr L1.{pos} + 1 = 2)
  \put(13,83){\vector(-1,-1){6}}
  \put(17,83){\line(1,-1){8}}
  \put(25,75){\vector(0,-1){17}}
\lCircle(7,\attr B2.{pos} = \attr L2.{pos} + 1 = 3)
  \put(5,72){\vector(0,-1){4}}
\lCircle(6,\attr B2.{val} = 2^{-\attr B2.{pos}} = 2^{-3})
\put(0,60){\line(1,0){115}}
\rCircle(5,\attr L3.{pos} = \attr L2.{pos} + 1 = 3)
  \put(25,52){\vector(0,-1){4}}
\rCircle(4,\attr B3.{pos} = \attr L3.{pos} = 3)
  \put(25,42){\vector(0,-1){4}}
\rCircle(3,\attr B3.{val} = 2^{-\attr B3.{pos}} = 2^{-3})
  \put(25,32){\vector(0,-1){4}}
\rCircle(2,\attr L3.{val} = \attr B3.{val})
  \put(23,23){\vector(-1,-1){6}}
  \put(5,62){\line(0,-1){37}}
  \put(5,25){\vector(1,-1){8}}
\cCircle(1,\attr L2.{val} = \attr L3.{val} + \attr B3.{val})
\end {picture}
\end{minipage}&
\raisebox{6mm}{
\begin{minipage}{5mm}
$s_1$
\par
\vspace*{28mm}
\par
$s_2$
\end{minipage}}
\end{tabular}

\end {center}
\caption {Slice partitioning}
\label{fig: program slice of AG}
\end {figure}

The \emph {dynamic program slice} of an execution of a statement $s$
in a program is a set of all the statements upon which $s$ depends,
directly or indirectly \cite {k90DynamicSlicing}.
In attribute grammars, attribute evaluation of a given parse tree can be 
considered to be
a sequence of evaluation of node attributes, and we can define dynamic program
slices for a given attribute grammar description and its input program.
For example, the sequence on the right side in Fig.~\ref{fig: program slice of
AG} is a possible attribute evaluation sequence of \attr L2.{val} for
AG in Fig.~\ref
{fig: grammar} and the input ``.011.''  The directed graph on the 
left side illustrates
direct dependency among the attributes.  The slice for a given attribute
(instance) is a set of attributes upon which the attribute
depends.\footnote{In AG, the slice can be given for an attribute instance 
rather than for evaluation of each attribute, because each attribute is 
evaluated only once.}  For example, the slice for \attr L3.{pos}
is a set of attributes upon which \attr L3.{pos} directly
or indirectly depends, namely, \attr L2.{pos} and \attr L1.{pos}.

\begin{figure*}[htb]
  \centerline{\includegraphics [width=0.9\linewidth]{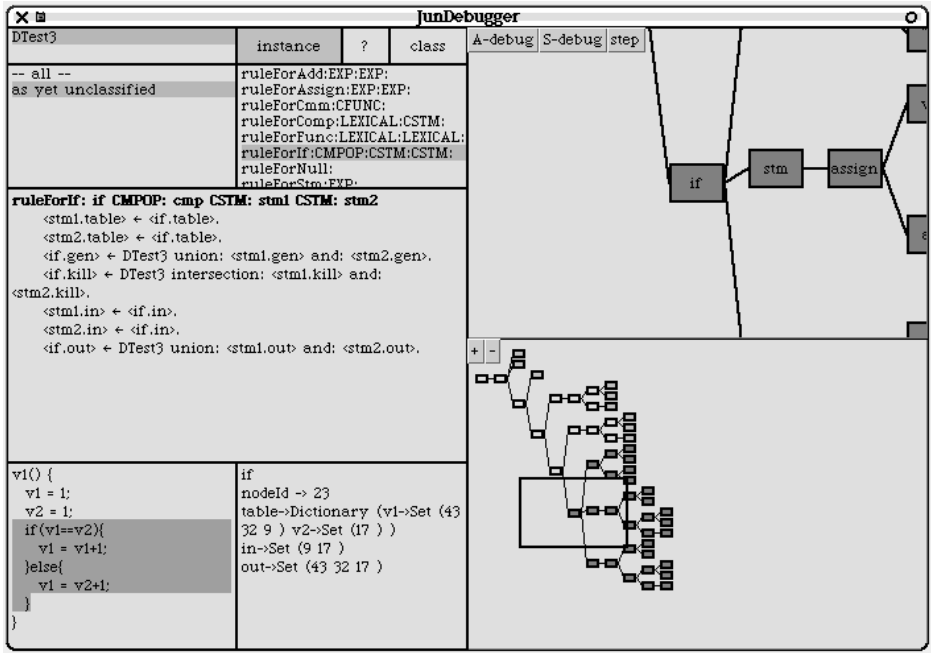}}
  \caption{Example of execution of Aki}
  \label{fig:Aki}
\end{figure*}

Our debugging strategy works as follows. Suppose that we already know
that a slice contains an incorrect attribution that is triggered by an
unknown bug.  The debugger partitions the attribute evaluation sequence
into arbitrary two subsequences (e.g., $s_1$ and $s_2$ in Fig.~\ref
{fig: program slice of AG}).  Then the debugger queries the user
about correctness of evaluation.  This is accomplished by asking
the correctness of the attribute values crossing the boundary between 
the two subsequences (e.g.,
correctness of \attr B2.{val} and \attr L2.{pos}).  If one of the 
crossing attribute values
turns out to be incorrect, then the debugger identifies that the error was 
triggered
in the subslice of that attribute.  Otherwise, when all the crossing attribute
values are correct, then the debugger excludes the subsequence $s_1$ from its
bug-locating search space.

For example in Fig.~\ref {fig: program slice of AG},
value assignment to \attr B2.{val}, $2^{-3}$ is different from its
expected value, $2^{-2}$ (\attr B2.{val} stands for
the value of binary number $(0.01)_2$).
Given this information from the user, the debugger understands that 
the bug inhabits somewhere in
the subslice of \attr B2.{val}.  
Next, the debugger divides the subslice of \attr B2.{val} and queries whether
the value of $\attr L2.{pos} = 2$ is correct.
The user answers "yes".
Then the debugger narrows the search space to \attr B2.{pos} and
\attr B2.{val},
and queries whether the value of $\attr B2.{pos} = 3$ is correct.
Since the user says it is incorrect, the debugger can locate the bug to
a single attribute \attr B2.{pos},
and it identifies that the attribution rule ``$\attr B2.{pos} = 
\attr L2.{pos} + 1$"
contains a bug.

\section{\aki : the debugger}
\label{sec: aki}

The \aki\ visual debugger is a part of our AG-based compiler
construction system that comprises a compiler frontend generator called
\rie\ and backend generator called \jun \cite {s91RieJun}.  \aki\ is used
to locate bugs in the compiler backend description.
The two systematic debugging techniques explained in earlier sections
are incorporated in \aki.  Because \jun\ accepts fairly large class of
attribute grammars, this approach can be applied to other AG evaluators
\cite {so96aki}.

When the compiler developer finds that the compiler generates an
incorrect code sequence for some source program, then he/she can supply
both the compiler backend description and the source program to \aki\ 
for debugging.  Fig.~\ref{fig:Aki} shows a screen shot of a debug
session using \aki.  The panes presents attribution rules, input source
program supplied to the debugged compiler, values of attributes, and
the parse tree of the source program in several forms.  These panes
work cooperatively:  user's interaction with one pane is reflected to
others.

\begin{figure}
  \centerline{\includegraphics [width=0.6\linewidth]{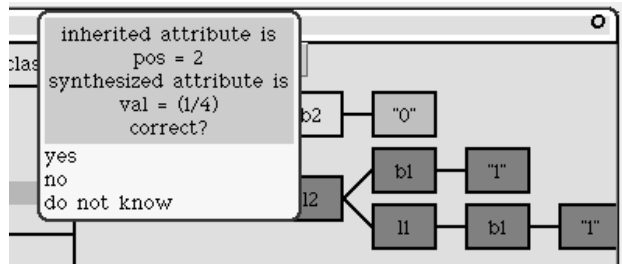}}
  \caption{Example of query of algorithmic debugging}
  \label{fig:query}
\end{figure}

\begin{figure}
  \centerline{\includegraphics [width=0.6\linewidth]{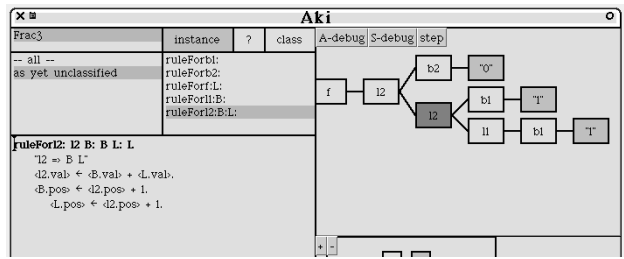}}
  \caption{The screen shot when the bug is located}
  \label{fig:bug}
\end{figure}

The user can choose systematic debugging strategies with buttons:
A-debug button for algorithmic debugging and S-debug button for
slice-based debugging.

The algorithmic debugger chooses an arbitrary node from the computation
tree and ask the user about correctness of the respective computation.
In Fig.~\ref {fig:query}, the debugger is asking about correctness of
the computed attribute value (``$\mathit {val} = 1/4$'').  \aki\ helps
the user answer this question by highlighting the respective subtree of
the parse tree in dark-gray and showing inherited attribute values
given as inputs to this computation (``$\mathit {pos} = 2$'').  The
user's answer to this question is used by the debugger to narrow the
search space for erroneous code in the program.  This narrowing is
repeatedly applied until the debugger eventually locates the erroneous
attribution rule (in Fig.~\ref {fig:bug}, \aki\ successfully located a
bug in \textsf {ruleForl2} rule.).

\begin{figure}
  \centerline{\includegraphics [width=0.6\linewidth]{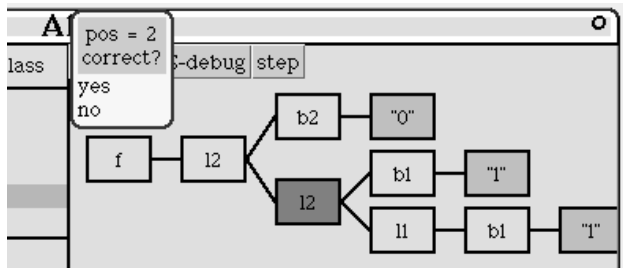}}
  \caption{Example of query of slice-based debugging}
  \label{fig:query-slice}
\end{figure}

The user starts slice-based debugger pointing it an incorrect attribute
value in the attributed parse tree.  The slice-based debugger
systematically locates the source of the problem in the attribution
rules.  The debugger computes the slice for the incorrect attribute and
partition the slice into two subslices.  Then the debugger queries if
incorrect attribute values are passed from one subslice to another.
This is accomplished by asking separate questions for these attribute
values, respectively.  In Fig.~\ref {fig:query-slice}, \aki\ is asking
about correctness of the value of one of those attributes.  Given
answers to these questions, the debugger narrows the search space for
erroneous code into one of the subslices and eventually locate the
incorrect attribution rule.

An advantage of slice-based debugging it can debug a program that
failed to complete in the middle of its execution.  On the other hand,
it is impossible to create a computation tree for the entire
computation of such program and hence difficult to apply algorithmic
debugging.

\section{Experiments and discussion}
\label{sec: discussion}

\subsection {Evaluation of \aki}

  \aki\ has been used to debug descriptions of several modules in our
complier construction project for the C language; examples of modules
are SSA-conversion and liveness analysis.  Users report it is easier to
find bugs by using \aki\ than by using conventional approach.

We have done user test to evaluate usefulness of \aki.  Three
experienced compiler programmers are chosen from our team as subjects.
They are shown a source program of some compiler module.  We have
included a typical mistake in the source program.  Then the programmer
is asked to locate using algorithmic debugging feature of \aki, using
slice-based debugging feature of \aki, or without support by \aki.  The
123 lines long test program contains 18 attribution rules.  Given a
sample C program that triggers the compiler's bug, the parser generates
a parse tree of 77 nodes and 171 attribute instances.

Fig.~\ref {fig: experiment} shows how the choice of debugging method
affects the time to locate bugs.  The numbers imply effectiveness of
the systematic debugging approach for AG-based compiler construction,
at least for smaller sized modules.

\begin{figure}
\begin{center}

\begin{tabular}{|c|c|c|c|}
\hline
           & algorithmic & slice-based & without \\
           & debugging   & debugging   & \aki    \\ \hline
number of  & 9           & 13          & NA      \\
queries    &             &             &         \\ \hline
time to    &             &             &         \\
locate bugs& 3           & 3           & 15      \\
(min.)     &             &             &         \\ \hline
\end{tabular}

\end{center}
\caption{Result of a user test}
\label{fig: experiment}
\end{figure}

\subsection {Discussion}

A known problem in algorithmic and automatic debugging is that the user
has to understand the question and answer it correctly (\cite {f92GADT}
\cite {Nilsson94} \cite {aadebug93}).

For example in the algorithmic debugging of AGs, the user may have
difficulties in checking whether the behavior of a Synth function or
the values of attributes are correct.  This is due to the fact that the
user has to look at the subtree in the parameter of the Synth function,
and furthermore attribute values may be a set and may have many
elements in compilers.

These problems have been partly solved from the algorithmic point of
view by extending the debugging algorithm in several situations [8].
For example, the extended algorithm can deal with the case when the
user find the value of the inherited attribute --- given as a premise of
a query by the system --- is itself wrong, or when the user cannot reply
to a query with confidence.  Some part of the extended algorithm is
implemented in \aki.

On the other hand, from the implementation point of view, we made
several efforts in making \aki\ so that the user can easily grasp the
attribute values and the subtree in a query.  For example, \aki\ can
show the source code corresponding to any subtree.  When a new
attribute value is computed by combining several attribute values with
some operation, \aki\ highlights the part of the original attribute
values within the new attribute value by a separate color, making the
difference of both attributes clear.

However, we think further improvement in the algorithm and the
implementation is necessary.

\section{Conclusion}
\label {sec: conclusion}

This paper has presented two systematic debugging techniques for
attribute grammars, one is based on the algorithmic debugging and the
other is based on the program slicing. These techniques have been
incorporated in our high-level visual debugging tool called \aki.

There remain some issues that require further investigation.  Currently
the two systematic debugging features are provided as separate
technologies.  We plan to integrate the two techniques and search for more
effective debugging methodology.

%\bibliographystyle{plain} 
%\bibliography{aadebug} 

\end{document}